\documentclass[12pt]{article}

\usepackage{times}

\usepackage[version=3]{mhchem}
\usepackage{authblk}
\usepackage{hyperref} 
\usepackage[capitalize,noabbrev]{cleveref}
\usepackage{booktabs}
\usepackage{longtable}
\usepackage{amssymb}
\usepackage[dvipsnames]{xcolor}
\usepackage{lineno}
\usepackage{setspace}

\usepackage[dvipdfmx]{graphicx}
\graphicspath{{Fig/}}

\usepackage[title]{appendix}

\usepackage{geometry}
\usepackage{pdflscape}
\usepackage{natbib}

\title{Accretion disk's magnetic field controlled the composition of the terrestrial planets}

\author[1,2,3]{William F. McDonough\thanks{Corresponding author. E-mail: mcdonoug@umd.edu}}
\author[2]{Takashi Yoshizaki}

\affil[1]{Department of Geology, University of Maryland, College Park, MD 20742, USA}
\affil[2]{Department of Earth Science, Graduate School of Science, Tohoku University, Sendai, Miyagi 980-8578, Japan}
\affil[3]{Research Center of Neutrino Sciences, Tohoku University, Sendai, Miyagi 980-8578, Japan}

\begin{document}

\maketitle

\doublespacing



\textbf{Chondrites, the building blocks of the terrestrial planets, have mass and atomic proportions of oxygen, iron, magnesium, and silicon totaling $\geq$90\% and variable Mg/Si ($\sim$25\%), Fe/Si (factor of $\geq$2), and Fe/O (factor of $\geq$3). The Earth and terrestrial planets (Mercury, Venus, and Mars) are differentiated into three layers: a metallic core, a silicate shell (mantle and crust), and a volatile envelope of gases, ices, and, for the Earth, liquid water. Each layer has different dominant elements (e.g., increasing Fe content with depth and increasing oxygen content to the surface). What remains an unknown is to what degree did physical processes during nebular disk accretion versus those during post-nebular disk accretion (e.g., impact erosion) influence these final bulk compositions. Here we predict terrestrial planet compositions and show that their core mass fractions and uncompressed densities correlate with their heliocentric distance, and follow a simple model of the magnetic field strength in the protoplanetary disk. Our model assesses the distribution of iron in terms of increasing oxidation state, aerodynamics, and a decreasing magnetic field strength outward from the Sun, leading to decreasing core size of the terrestrial planets with radial distance. This distribution would enhance habitability in our solar system, and would be equally applicable to exo-planetary systems.}


The formation of metallic cores in terrestrial planets greatly influences the thermal and biological evolution of a planet. Core formation concentrates the heat producing elements (i.e., potassium, thorium and uranium) into the insulating, outer silicate shell and produces a conductive fluid, which can create a planetary magnetic field. The mass fraction of metallic core in Mercury, Venus, Earth, and Mars decreases with heliocentric distance from about 2/3, to 1/3 (Venus and Earth), to 1/5, respectively \citep{sohl2015interior}. What chemical and/or physical process produced the large variation observed in Fe/O values in chondrites and the terrestrial planets, particularly for Mercury? The presence of a long-lived, internally convecting metallic core results in dynamo action and the generation of a planet's surrounding protective magnetosphere that nurtures life. These differentiated planets represent the most likely home for life and its evolution. \bigskip

The compositions of the terrestrial planets and chondritic asteroids record, on average, an outward increase in oxygen fugacity, potentially a decrease in the nebular condensation temperature, and decreasing amounts of metallic iron contributing to planet building. The redox and temperature gradients leads to less metallic iron and more H- and O-rich solids (i.e., phyllosilicates) outward in the solar system. Importantly, nebular condensates do not reach the high Fe/Si values of Mercury even with strongly reduced, high-temperature conditions \citep{ebel2011equilibrium}. Thus, further metal-oxide separation processes are needed. \bigskip

 Compositional models for the terrestrial planets are constructed from the following data sets: composition of the Sun (i.e., $ > $99\% mass of the solar system), chemical trends for samples from a planet, satellite observations, and compositions of chondritic meteorites (i.e., the solar system's building blocks of undifferentiated rock and metal mixtures). Importantly, the chondrites that we have, however, are those leftover from planet building.  Here we use our earlier compositional models for the Earth \citep{mcdonough2014compositional} and Mars \citep{yoshizaki2019mars_long} and model the recent data from the MESSENGER mission to Mercury to predict its bulk composition (\cref{tab:composition}), which is consistent with known physical and chemical constraints (Methods). Given limited data for Venus, which is consistent with an Earth-like analog \citep{surkov1987uranium,dumoulin2017tidal}, we assume it has a bulk Earth composition.  \bigskip

\begin{figure}[p]
\centering
\includegraphics[width=.7\linewidth]{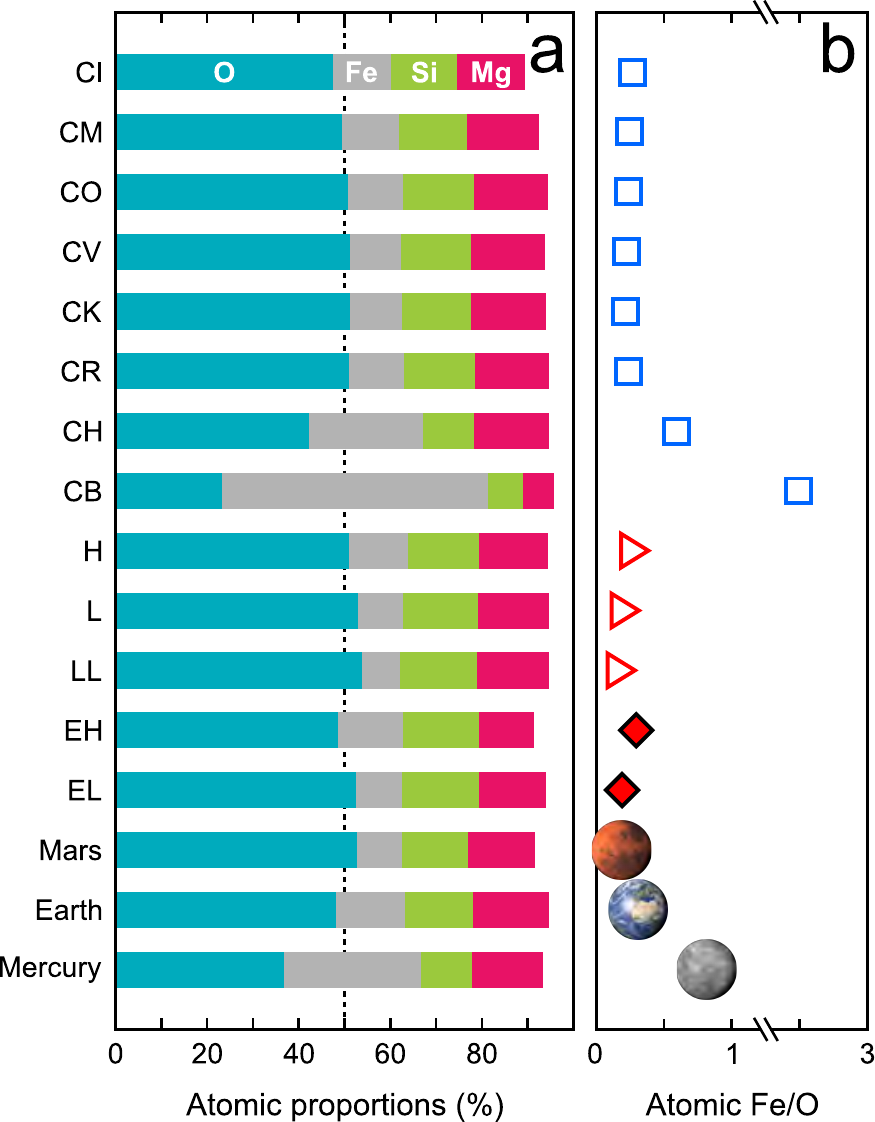}
\caption{\textbf{(a) Atomic abundances of major elements and (b) Fe/O values in the solar system bodies.} Data sources are given in supplementary materials.}
\label{fig:M_elements}
\end{figure}

\begin{figure}[p]
\centering
\includegraphics[width=1\linewidth]{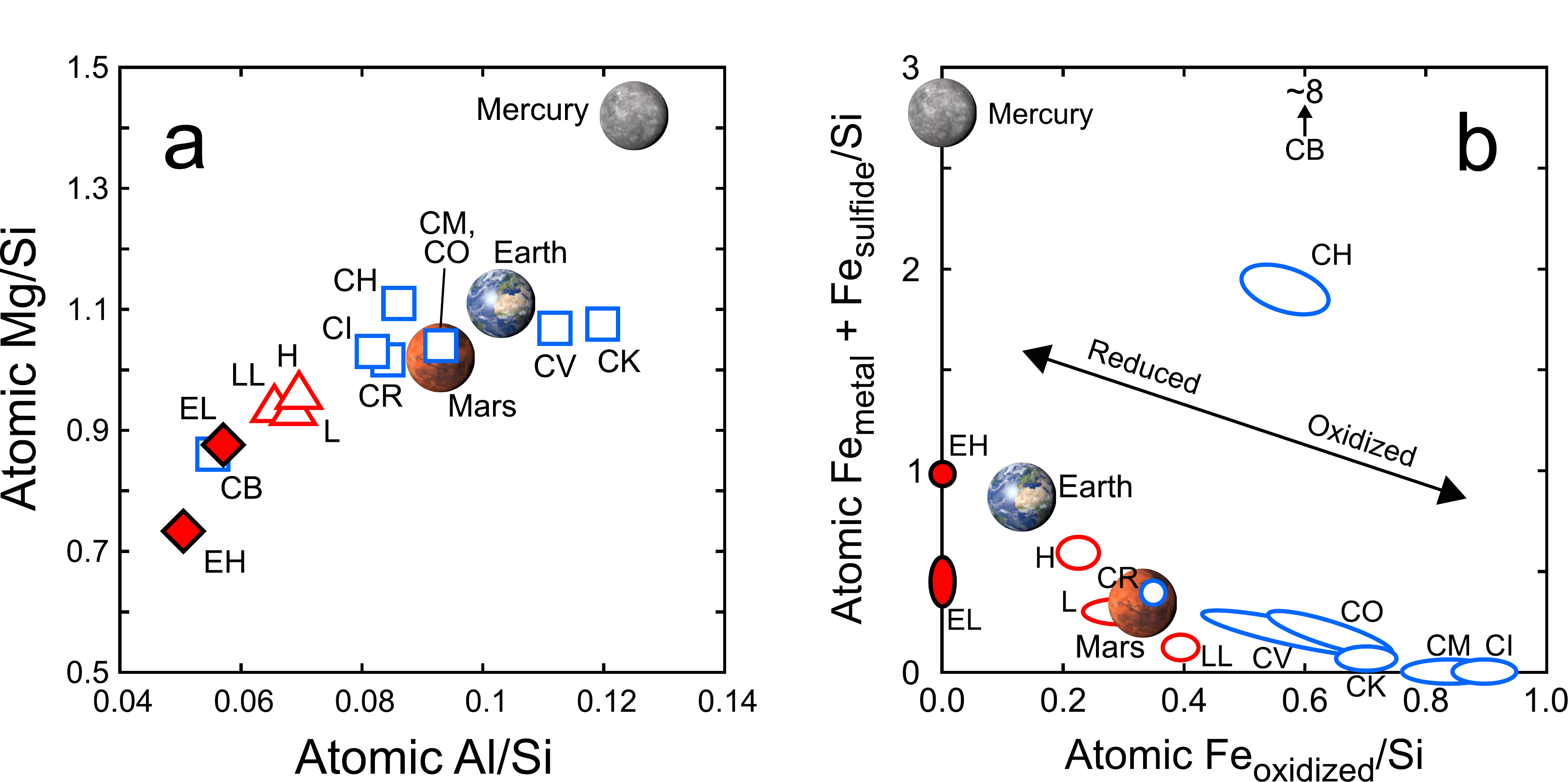}

\caption{\textbf{Ratios of major cations in the terrestrial planets and chondrites.} (a) Magnesium/Si versus Al/Si. (b) Abundances of reduced (metal and sulfide) and oxidized Fe normalized to Si. Data sources are given in supplementary materials. Red symbols identify the inner solar system, NC chondrites; see text for details.}
\label{fig:UC_Al-Si}
\end{figure}

Chondrites are geologically unprocessed materials, with their chemical compositions reflecting the local nebular conditions. However, chondrites differ markedly in their redox states and major element compositions (\cref{fig:M_elements,fig:UC_Al-Si}). The redox state, mineralogies, and isotopic compositions of chondrites and other meteorites demonstrate that the early solar system was not compositionally homogeneous \citep{warren2011stable}. Significantly, the less oxidized Non-Carbonaceous (NC) meteorites, including the enstatite and ordinary chondrites (\cref{fig:UC_Al-Si}), are viewed as coming from inner solar system regions closer to the Sun (i.e., mostly 2 to 3 AU) than the oxidized Carbonaceous Chondrites (CC) and related meteorites \citep{kruijer2017early}. The redox state and a dozen or so isotopic systems now link enstatite chondrites and Earth and equally, ordinary chondrites and Mars \citep{warren2011stable}. \bigskip

The Fe content of chondrites typically ranges from 1/5 to $<$1/3 of their total mass, with atomic Fe/Si typically 0.74\;$\pm$\;0.12, whereas some rare chondrite groups (i.e., CB, CH, and G types) can have Fe/Si values up to 8 (\cref{fig:UC_Al-Si}b). The amount of iron accreted by a chondritic or terrestrial planet is not set by any particular rule; in the nebula, condensing iron-nickel grains, and other metallic alloys, are distinctly influenced by aerodynamic, gravitational, photophoretic, and electromagnetic sorting forces, compared to silicates \citep{weidenschilling1978iron,wurm2013photophoretic,kruss2018seeding}. \bigskip

A planet's mass fraction of core to silicate shell reflects the accretion disk's average local redox state, accretion proportions of metal relative to silicates, and the redox conditions accompanying core-mantle differentiation. Impact-induced erosion/evaporation can also modify the mass fraction of core. Accretion sets the planetary values of Fe/O, Fe/Si, and Fe/Mg, which accounts for $\sim$93\% of its mass, with the addition of the minor elements, Ca, Al, Ni, and S, bringing the total to 99 wt\% (\cref{fig:M_elements}). A planet's mantle Mg\# (i.e., atomic Mg/(Mg+Fe)), therefore, is the recorder of the average redox state that accompanied core-mantle differentiation. \bigskip

Mercury has the largest metallic core, high Fe/Al (21.8) and Mg/Si (1.4) values (\cref{tab:composition,fig:UC_Al-Si}) and a silicate sphere with negligible Fe (i.e., Mg\# $\sim$ 0.99) \citep{nittler2017chemical}. By comparison, Mars has a smaller metallic core, a lower atomic Fe/Al (7.3) and Mg/Si (1.02) and a silicate shell with a low Mg\# (0.79) than Earth (9.7, 1.11, and 0.89, respectively) (\cref{tab:composition}). Therefore, the silicate shells of Mars, Earth (and Venus), and Mercury get progressively relatively smaller, their cores relatively bigger, and their mantles more Mg\#-rich with heliocentric distance. \bigskip

\begin{figure}[p]
\centering
\includegraphics[width=1\linewidth]{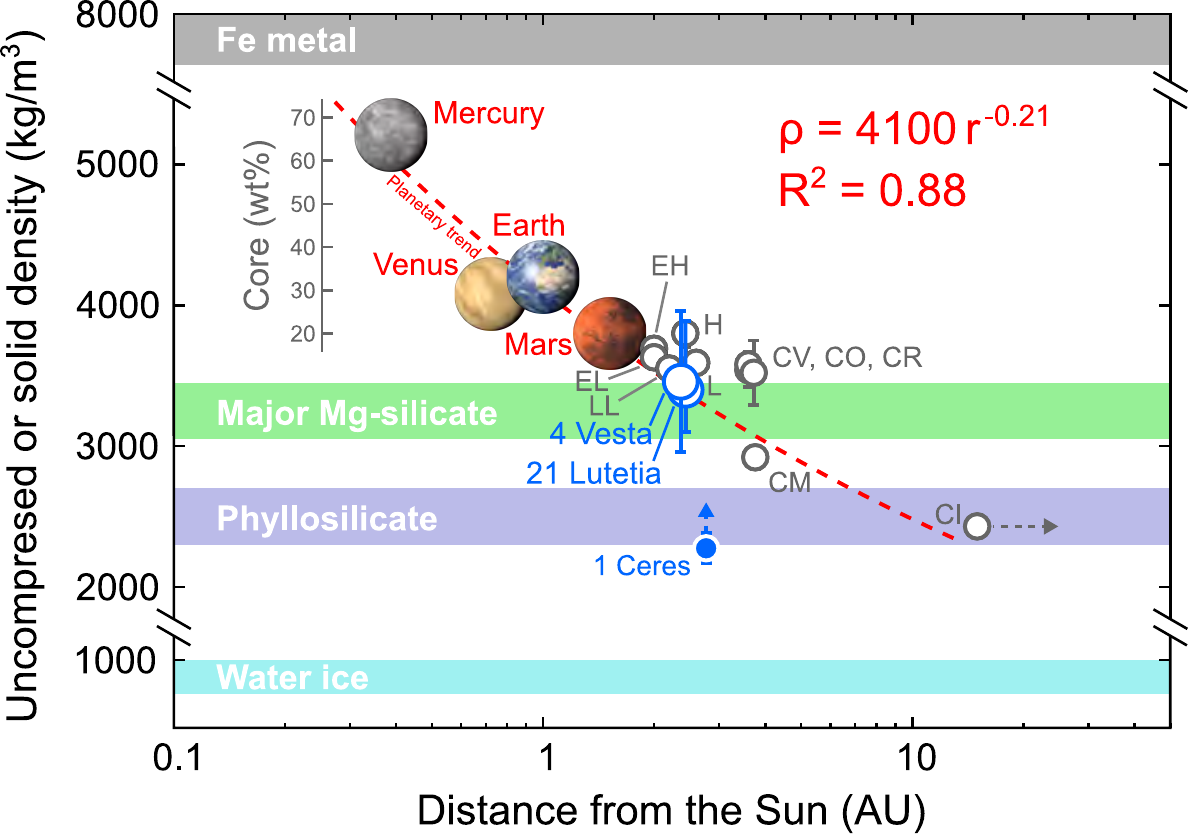}
\caption{\textbf{Density of the solar system bodies.} Uncompressed and solid densities are shown for terrestrial planets and chondrites (grey), respectively. Bulk planetary densities are shown for asteroids (blue). For 1 Ceres, its bulk density is a lower limit of its solid density, given its high ice abundance and porosity. The red line shows a fit curve for the planets ($\rho = 4,100 r^{-0.21}$). Data sources are given in supplementary materials.}
\label{fig:density_vs_r}
\end{figure}

Importantly, we show that the mean atomic number of the terrestrial planets increases inwards and show a correlation between heliocentric distance and a planet's uncompressed density (\cref{fig:density_vs_r}). This planetary density trend also extends to bodies in the asteroidal belt: undifferentiated and differentiated asteroids are less dense than the terrestrial planets (\cref{fig:density_vs_r}). Most chondrites have sub-solar Fe/Si (\cref{fig:UC_Al-Si}b) and show a marked variation in Fe/Si and metallic iron content. Thus, the outward decrease of planetary density appears to primarily reflect dynamic metal-oxide separation in the protoplanetary disk, rather than post-accretionary processes. \bigskip

Many processes can lead to chemical segregation in a protoplanetary disk, including aerodynamic sorting of metal and oxide \citep{weidenschilling1978iron}, and electromagnetic separation of magnetized micro-particles of Fe-Ni alloys (existing below their Curie point with a stable magnetic spin) from silicates \citep{harris1967fractionation,larimer1970chemical}. Metal-oxide separation during chondrule formation might also be an effective separation process \citep{connolly2001formation}. Recently, Wurm et al. \citep{wurm2013photophoretic} also observed that photophoretic separation enhances metal-silicate fractionation. Thus, various physical sorting mechanisms in the nebular likely established the variation in Fe/O, Fe/Si, and Fe/Mg in chondrites and terrestrial planets.  The degree of inward enrichment of magnetic micro-particles in the protoplanetary disk appears to scale radially with its magnetic field strength.  \bigskip

We show a correlation between the metallic core fraction and heliocentric distance for the terrestrial planets (\cref{fig:density_vs_r}). This trend is interpreted as resulting from a decay of magnetic field strength in the protoplanetary disk and is consistent with paleomagnetic intensities recorded in meteorites \citep{bryson2020constraints,fu2020weak}, magneto-hydrodynamic (MHD) disk models \citep{bai2015hall}, and a simple Biot-Savart law that describes a 3D radial dependent magnetic field strength:
\begin{equation} \label{Biot-Savart}
B = \frac{\mu_0 I}{2 \pi r} 
\end{equation}
\noindent where $B$ is magnetic flux density (Tesla, or kg s$^{-2}$ A$^{-1}$), $\mu_0$ is the magnetic constant (H m$^{-1}$ or N A$^{-2}$), $I$ is current (A), $r$ is position in 3D-space (m) (Methods and Supplementary \cref{fig:B-S_plot}). Azimuthal magnetic fields are likely to be strongest in the midplane \citep{wardle2007magnetic}, the site of planetary accretion. The magnetic field strength of protoplanetary disk is recognized as a constantly evolving 4 dimensional phenomena that have been extensively treated by sophisticated MHD modeling (e.g., \citep{wardle2007magnetic}). Our perspective simply takes as a set of average conditions for the midplane of the disk at 4 critical distances: Mercury (500 $\mu$T, assumes a saturation field \citep{levy1978magnetic}), Earth (100 $\mu$T \citep{wardle2007magnetic}), and Vesta ($\sim$50 $\mu$T \citep{wang2017lifetime}), with interpolations for Mars (60 $\mu$T). This simple model (\cref{fig:B-S_plot}) appeals to a fundamental scaling relationship ($B \propto r^{-1.3}$) that views the magnetic field strength at the disk's midplane decreasing with distance from the central current moment of the evolving Sun. MHD modeling predicts $ B \propto r^{-1.6}$ \citep{wardle2007magnetic,bai2015hall}. With mT magnetic field strength of the disk at $\sim$1 AU is approximately sub-equal in its horizontal and vertical components and increases significantly towards the disk's center \citep{wardle2007magnetic}. Kruss and Wurm \citep{kruss2018seeding} suggested the enhanced growth of iron-rich planetesimals in the inner region of protoplanetary disks leads to an iron gradient in the solar system. A corollary of this model is that planetary migration in the inner solar system did not disrupt this final result and that the Fe content can help us figure out where an object formed.  \bigskip

Nebular gas carries the magnetic force in the protoplanetary disk. Metal-silicate segregation by solar magnetic forces is only effective before nebular gas dissipation \citep{kruss2020composition,wang2017lifetime}. Mercury's formation under reduced conditions in the presence of a gas disk with a strong magnetic field implies a $\tau_{accretion}$ age that is restricted to when a nebular envelope existed. Accretion in the presence of nebular gas is consistent with pebble accretion models of planetary formation, which may have contributed to the early formation of Mars and the iron meteorite parent bodies \citep{johansen2015growth}. Therefore, a key implication of our model is an early formation age ($\sim$2 Myr) for Mercury, comparable to that of Mars \citep{dauphas2011perspective,yoshizaki2020earth}. Furthermore, our model does not require a giant impact event and partial loss of its silicate shell for the origin of Mercury's large core  \citep{benz2007origin,asphaug2014mercury}. \bigskip

Incorporation of silicon, sulfur, and other elements into the Fe-Ni core of Mercury plays a minor role in establishing its size. Our model does not identify Mercury's core composition; it remains a considerable unknown. We estimate a modest sulfur content for Mercury (2.5 $\pm$ 0.5 wt\% (Methods) (cf., \citep{namur2016sulfur,boujibar2019u}) based on Mercury's high bulk K/Th (6,000 to 8,000) \citep{nittler2017chemical}. The presence of silicon, sulfur, and other light elements, however, correspondingly lowers the solidus of the core and keeps it molten and convecting, thus providing the critically important conditions for dynamo action and magnetosphere generation. \bigskip

The presence and size of a planetary metallic core depend on the redox environment of the protoplanetary disk and the planet's mass fraction of accreted metal alloys, the latter of which is controlled by electromagnetic processing in the disk. A planet's volatile element inventory (e.g., bulk sulfur content), coupled with the redox condition of core formation, controls the amount of light element in the core, which significantly lowers the core's solidus and extends its molten-state lifetime. Collectively, these factors contribute to convection in a molten core and dynamo action. These attributes of our solar system would be equally applicable to exo-planetary systems. The generation of a planetary magnetosphere, which nurtures life, shapes a planet's habitability. It is likely that life's sustainability critically depends on being sited in the Goldilocks zone and having the right amount of metallic core, which contains an appropriate of a light element and is not cooling too fast.  \bigskip

\section*{Methods}
\label{sec:methods}

\subsection*{Model description}

The bulk composition of Mercury was defined to be consistent with its geodetic observables: mass, density, and MOI (moment of inertia) \citep{margot2018mercury}. Assuming a thermal gradient in the protoplanetary disk and the higher condensation temperature for forsterite (\ce{Mg2SiO4}) versus enstatite (\ce{Mg2Si2O6}), then we predict a gradient in atomic Mg/Si (1.4) and Al/Si (0.12) values projecting from the values for Mars and Earth (\cref{fig:UC_Al-Si}). The assumed Fe/Si (2.7) was established from a mass fraction of silicate to metal to be consistent with Mercury's uncompressed density (\cref{fig:density_vs_r}), assuming proportions (40:53:7) and densities of silicates, metals, and sulfides (3,100, 7,100, and 4,600 kg m$^{-3}$, respectively) as nominal values and with the metal alloy being a Fe-Ni-Si mixture. It is also likely that core-mantle differentiation add some fraction of silicon into the core under reducing conditions \citep{nittler2017chemical}. The atomic Ca/Al (0.73) and Fe/Ni (18.3) values are those seen in chondrites \citep{alexander2019quantitative_CC,alexander2019quantitative_NC}. The sulfur content was based on the planetary volatility trend established from Mercury's measured K/Th value and an extrapolation to the condensation temperature, following the practice used in \citep{mcdonough1995composition,yoshizaki2019mars_long}.
Ratios of alkali metals to refractory elements provide a constraint on a planet's volatile element depletion trend, however, this simple model can be influenced by redox conditions and core formational processes. Finally, the oxygen content was set to make the total equal to 100\%. This model establishes the bulk properties of Mercury, but it does not specify the distribution of elements between the core and mantle. \bigskip

\subsection*{Data sources}

Data sources for \cref{fig:M_elements,fig:UC_Al-Si} are as following: chondrites \citep{urey1953composition,alexander2019quantitative_CC,alexander2019quantitative_NC,mccall1968bencubbin,ivanova2008isheyevo,gosselin1990chemical,wasson1990allan,bischoff1993acfer}; Mercury (this study); Earth \citep{mcdonough2014compositional}; Mars \citep{yoshizaki2019mars_long}. For \cref{fig:density_vs_r}, density of planetary bodies are from \citep{russell2012dawn,park2019high,sierks2011images,consolmagno2006density,macke2010enstatite,macke2011density,britt2003stony,lewis1972metal,stacey2005high}; heliocentric distances of chondrite parent bodies are from \citep{desch2017effect}.

\bibliographystyle{elsarticle-harv}
\bibliography{myrefs}

\section*{Acknowledgments}
We thank many our colleagues who have listened to various versions of this project and given helpful comments. WFM gratefully acknowledges NSF support (EAR1650365). TY acknowledges supports from the Japanese Society for the Promotion of Science (JP18J20708) and the GP-EES and DIARE programs.

\section*{Author contributions}
WFM and TY proposed and conceived various portions of this study and together calculated the compositional models. The manuscript was written by WFM, with edits, discussions, and revisions by TY \& WFM. Both authors read and approved the final manuscript.

\section*{Competing interests}

The authors declare no competing interests.

\section*{Data and materials availability}

Correspondence and requests for materials should be addressed to WFM.

\section*{Supplementary information}

Supplementary information is available for this paper.

\clearpage

\begin{landscape}
\begin{table}[p]
\centering
\caption{\textbf{Composition of the terrestrial planets.}}
\label{tab:composition}
\begin{tabular}{ccccc}
\toprule
    Atomic\% & Mercury & Earth \& Venus   & Mars  & CI chondrite   \\
          & (see Methods) &  \citep{mcdonough2014compositional}  & \citep{yoshizaki2019mars_long}  &(volatile-free) \\
          & & & & \citep{alexander2019quantitative_CC}  \\
\midrule
    O     & 36.8  & 49.0  & 55.3 & 48.2 \\
    Mg    & 15.6  & 16.7  & 15.3 & 15.1 \\
    Si    & 11.0  & 15.1  & 15.1 & 14.7 \\
    Fe    & 30.0  & 15.1  & 10.3 & 12.8 \\
    Ni    & 1.64  & 0.82  & 0.57 & 0.70 \\
    Al    & 1.38  & 1.56  & 1.41 & 1.20 \\
    Ca    & 1.00  & 1.13  & 1.03 & 0.88 \\
    S     & 2.49  & 0.52  & 0.92 & 6.43 \\
          &       &       &  \\
    Fe/Si & 2.72  & 1.00  & 0.69 & 0.87 \\
    Fe/Al & 21.8  & 9.7   & 7.3 & 10.6 \\
    Fe/O  & 0.82  & 0.31  & 0.19 & 0.26 \\
    Mg/Si & 1.42  & 1.11  & 1.02 & 1.03 \\
    Al/Si & 0.12  & 0.10  & 0.09 & 0.08 \\
          &       &       &       &  \\
    Mean \textit{Z} & 15.4 & 12.7 & 11.8 & 12.6 \\
\midrule
Mass\%   & Mercury & Earth \& Venus & Mars & CI (volatile-free)  \\ 
\midrule
O      & 18.3   & 29.7  & 36.3 & 29.1  \\
Mg     & 11.8   & 15.4  & 15.3 & 13.9  \\
Si     & 9.65   & 16.1  & 17.4 & 15.6  \\
Fe     & 52.3   & 32.0  & 27.3 & 26.9  \\
Ni     & 3.00   & 1.82  & 1.36 & 1.54  \\
Al     & 1.16   & 1.59  & 1.56 & 1.22  \\
Ca     & 1.25   & 1.71  & 1.69 & 1.33  \\
S      & 2.50   & 0.64 & 1.21 & 7.78  \\
\bottomrule
\end{tabular}
\end{table}
\end{landscape}

\clearpage

\clearpage

\clearpage

\begin{appendices}

\section*{Supplementary information}

\setcounter{figure}{0}
\setcounter{table}{0}
\renewcommand{\figurename}{Supplementary Figure}

\begin{figure}[h]
\centering
\includegraphics[width=1\linewidth]{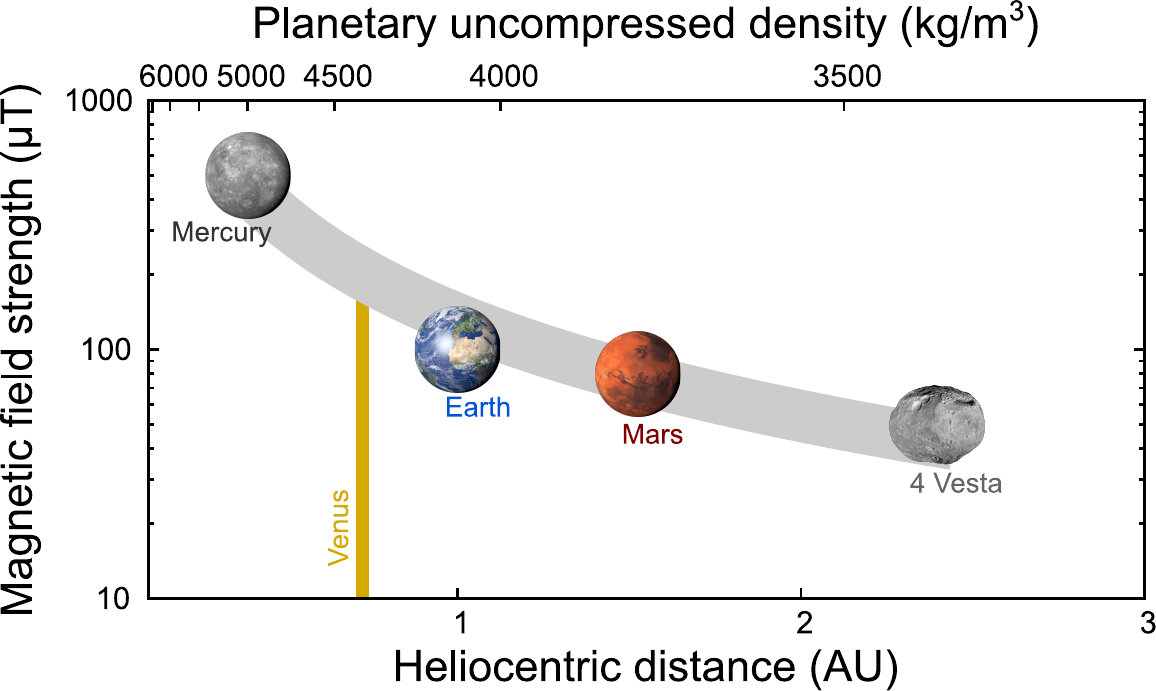}
\caption{A simple Biot-Savart type model for the average magnetic field strength versus accretion position of the terrestrial planets in the protoplanetary disk. A top x-axis shows the correlation between uncompressed density and heliocentric distance. See text for details and data. A prediction for Venus is shown with a bar; its uncompressed density is 4,100 kg m$ ^{-3}$  \citep{lewis1972metal,stacey2005high}.}
\label{fig:B-S_plot}
\end{figure}
\clearpage

\end{appendices}

\end{document}